\documentclass[%
 reprint,
 amsmath,amssymb,
 aps, nofootinbib
]{revtex4-1}
\usepackage{graphicx}
\usepackage{dcolumn}
\usepackage{bm}

\usepackage{amsmath,amssymb}
\usepackage{slashed}
\usepackage{graphicx}
\usepackage{dsfont}
\usepackage{bm}
\usepackage[top=1.0in,bottom=1.0in,left=1.0in,right=1.0in]{geometry}
\usepackage[colorlinks,linkcolor=blue,citecolor=blue]{hyperref}
\usepackage{CJKutf8}
\usepackage[caption=false]{subfig}
\usepackage{booktabs} 
\usepackage{ulem}

\begin{document}
\setlength{\oddsidemargin}{0.5cm}
\setlength{\topmargin}{-0.1cm}
\setlength{\textheight}{21cm}
\setlength{\textwidth}{15cm}
\newcommand{\be}{\begin{equation}}
\newcommand{\ee}{\end{equation}}
\newcommand{\bea}{\begin{eqnarray}}
\newcommand{\eea}{\end{eqnarray}}
\newcommand{\ba}{\begin{eqnarray}}
\newcommand{\ea}{\end{eqnarray}}

\newcommand{\fslash}{\hspace{-1.4ex}/\hspace{0.6ex} }
\newcommand{\Dslash}{D\hspace{-1.6ex}/\hspace{0.6ex} }
\newcommand{\Wslash}{W\hspace{-1.6ex}/\hspace{0.6ex} }
\newcommand{\pslash}{p\hspace{-1.ex}/\hspace{0.6ex} }
\newcommand{\kslash}{k\hspace{-1.ex}/\hspace{0.6ex} }
\newcommand{\underkslash}{{\underline k}\hspace{-1.ex}/\hspace{0.6ex} }
\newcommand{\epslash}{{\epsilon\hspace{-1.ex}/\hspace{0.6ex}}}
\newcommand{\partslash}{\partial\hspace{-1.6ex}/\hspace{0.6ex} }

\newcommand{\nn}{\nonumber}
\newcommand{\Tr}{\mbox{Tr}\;}
\newcommand{\tr}{\mbox{tr}\;}
\newcommand{\ket}[1]{\left|#1\right\rangle}
\newcommand{\bra}[1]{\left\langle#1\right|}
\newcommand{\rhoraket}[3]{\langle#1|#2|#3\rangle}
\newcommand{\brkt}[2]{\langle#1|#2\rangle}
\newcommand{\pdif}[2]{\frac{\partial #1}{\partial #2}}
\newcommand{\pndif}[3]{\frac{\partial^#1 #2}{\partial #3^#1}}
\newcommand{\pbm}[1]{\protect{\bm{#1}}}
\newcommand{\avg}[1]{\left\langle #1\right\rangle}
\newcommand{\vnabla}{\mathbf{\nabla}}
\newcommand{\notes}[1]{\fbox{\parbox{\columnwidth}{#1}}}
\newcommand{\pair}{\raisebox{-7pt}{\includegraphics[height=20pt]{pair0.pdf}}}
\newcommand{\paircrs}{\raisebox{-7pt}{\includegraphics[height=20pt]{pair0cross.pdf}}}
\newcommand{\paircc}{\raisebox{-7pt}{\includegraphics[height=20pt]{pair0cc.pdf}}}
\newcommand{\paircrscc}{\raisebox{-7pt}{\includegraphics[height=20pt]{pair0crosscc.pdf}}}
\newcommand{\pairloop}{\raisebox{-7pt}{\includegraphics[height=20pt]{pairloop.pdf}}}
\newcommand{\pairloopf}{\raisebox{-7pt}{\includegraphics[height=20pt]{pairloop4.pdf}}}
\newcommand{\pairlooph}{\raisebox{-7pt}{\includegraphics[height=20pt]{pair2looph.pdf}}}

\newcommand{\YZ}[1]{{\color{blue}YZ:{#1}}}


\title{Collins-Soper Kernel in the QCD Instanton Vacuum}

\author{Wei-Yang Liu}
\email{wei-yang.liu@stonybrook.edu}

\author{Ismail Zahed }
\email{ismail.zahed@stonybrook.edu}
\affiliation{Center for Nuclear Theory, Department of Physics and Astronomy, Stony Brook University, Stony Brook, New York 11794-3800, USA}

\author{Yong Zhao}
\email{yong.zhao@anl.gov}
\affiliation{Physics Division, Argonne National Laboratory, Lemont, IL 60439, USA}

\date{\today}
\begin{abstract}
We outline a general framework for evaluating the non-perturbative soft functions in the QCD instanton vacuum. In particular, from the soft function we derive the Collins-Soper (CS) kernel, which drives the rapidity evolution of the transverse-momentum-dependent parton distributions. The resulting CS kernel, when supplemented with the perturbative contribution,
agrees well with recent lattice results and some phenomenological parametrizations. Moreover, our CS kernel depends logarithmically on the large quark transverse separation, providing a key constraint on its phenomenological parametrization. Finally, a lattice calculation can be directly compared to our generic results in Euclidean signature, thus providing a new approach for evalulating the soft function and extracting the CS kernel by analytical continuation.
\end{abstract}

\maketitle

\section{Introduction}
The transverse-momentum-dependent parton distributions (TMDs) describe the distribution of 
partons in the transverse plane of a fast moving hadron, both in longitudinal momentum $x$ and transverse position $b_T$, for fixed resolution $\mu$ and at a given rapidity scale $\zeta$, i.e., 
$f(x,b_T,\mu,\zeta)$. Modulo a sign, TMDs are
used to describe Drell-Yan (DY) and semi-inclusive deep inelastic processes (SIDIS). With the help of factorization in quantum chromodynamics (QCD)~\cite{Collins:1981uk,Collins:1981va,Collins:1984kg}, the latter are used to empirically extract TMDs through global fitting~\cite{Davies:1984sp,Ladinsky:1993zn,Landry:2002ix,Konychev:2005iy,Sun:2014dqm,DAlesio:2014mrz,Bacchetta:2017gcc,Scimemi:2017etj,Bertone:2019nxa,Scimemi:2019cmh,Bacchetta:2019sam,Hautmann:2020cyp,Bury:2022czx,Bacchetta:2022awv,Moos:2023yfa,Isaacson:2023iui,Aslan:2024nqg,Bacchetta:2024qre,Yang:2024drd,Bacchetta:2024yzl,Kang:2024dja}. 

The TMDs are intrinsically non-perturbative. Defined from parton correlation functions on the light front, the TMDs also include ultraviolet (UV) and the so-called rapidity divergences. After renormalization, their rapidity evolution is driven by the Collins-Soper (CS) kernel~\cite{Collins:1981uk},  which depends on the parton transverse momentum, and becomes non-perturbative in the infrared (IR) region.
In recent years, thanks to the proposal of large-momentum effective theory~\cite{Ji:2013dva,Ji:2014gla,Ji:2020ect}, the TMDs and the CS kernel can be calculated from the quasi-TMDs~\cite{Ji:2014hxa,Ji:2018hvs,Ebert:2018gzl,Ebert:2019okf,Ji:2019sxk,Ji:2019ewn,Ji:2021znw,Ebert:2022fmh,Zhao:2023ptv} in lattice QCD. The lattice extraction of the CS kernel is especially performed by quasi-TMDs calculation at different target hadron momenta~\cite{Shanahan:2020zxr,Shanahan:2021tst,LatticeParton:2020uhz,Li:2021wvl,LatticePartonLPC:2022eev,Shu:2023cot,LatticePartonLPC:2023pdv,Avkhadiev:2023poz,Avkhadiev:2024mgd,Bollweg:2024zet}.  This first-principles extraction of the CS kernel, when combined with global analysis, can enhance
considerably our description of the partonic content of a hadron~\cite{Billis:2024dqq}.

At low resolution, the QCD vacuum in Euclidean signature is  populated with topologically active gauge fields, which interpolate between vacua of different Chern-Simons numbers~\cite{Leinweber:1999cw}. Numerical evidence follows from
dedicated QCD lattice configurations after the use of the gradient flow technique~\cite{Michael:1994uu,Michael:1995br,Bakas:2010by,Athenodorou:2018jwu}. In the instanton liquid model (ILM), these transitions are well described by pseudo-particles or instantons and anti-instantons, localized and strong  gauge fields with a typical size of $1/3\,\rm fm$, that tunnel  at a rate of about $1/\rm fm^4$. Thorough descriptions of the ILM can be found in~\cite{Shuryak:1997vd,Schafer:1996wv,Shuryak:1992jz,Shuryak:1992ke,Schafer:1993ra} (and references therein).

The purpose of this work is to present a calculation of the CS kernel in the ILM of the QCD vacuum. In Section~\ref{SECII} we briefly outline the general features of the Wilson loops used to construct the pertinent soft functions for DY, SIDIS and $e^+e^-$ pair annihilation in Minkowski signature. In Section~\ref{SECIII} we introduce the concept of 
the soft functions in the Collins (off-the-light-cone) scheme~\cite{Collins:2011zzd}, that are amenable to an analysis in the ILM in Euclidean signature. Key to this concept is the notion of Wilson loops on contours with rapidity angles $\theta$ that analytically continue to rapidity in Minkowski signature. In Section ~\ref{SECIV} we detail the result for the CS kernel
as extracted from the ILM and its evolution. Our conclusions are in Section~\ref{SECV}. Further details in support of the derivations, are presented in the Appendices.

\section{Wilson loops and soft functions}
\label{SECII}
Wilson lines are important ingredients in QCD factorization, as they enforce gauge invariance of bi-local operators, in the twist expansion of pertinent correlators on the light cone~\cite{Collins:2011zzd}. 
They resum the gluon emission between the bi-local sources, and 
also compose the building parts of the above mentioned Wilson loops
in Minkowski signature. Generally, a straight Wilson line from 
point $x$ to  point $y$ is defined as
\begin{equation}
    W[x;y]=\mathcal{P}\exp\left[ig\int_{x}^{y} dz_\mu A^\mu(z)\right]\,.
\end{equation}

\subsection{Wilson loops}
Wilson loops  made of Wilson lines capture important aspects of non-perturbative physics in QCD. In Euclidean signature, they are central for our understanding of IR physics in QCD, such as confinement in the vacuum~\cite{Diakonov:2009jq,Simonov:1996ati}, deconfinement in matter~\cite{DeMartini:2021xkg,Zhitnitsky:2006sr}, and the spin forces between heavy quarks ~\cite{Shuryak:2021hng,Shuryak:2021fsu}. A full formulation of gauge theories in loop space
was also pursued in~\cite{Makeenko:1980vm,Makeenko:1979pb,Polyakov:1987ez}.
In Minkowski signature, Wilson loops can be used to define the forward scattering amplitudes of selected hadronic processes in the eikonal approximation~\cite{Korchemskaya:1994qp}, and appear in 
a number of inclusive and semi-inclusive processes in high energy scattering using factorization~\cite{Cherednikov:2020mtu,Cherednikov:2017qbt}. 

For smooth and simple contours $C$, the generic Wilson loop is
\begin{equation}
\label{loop}
    W(C)=\frac1{N_c}\mathrm{Tr}\left\langle \mathcal{P} \exp\left(ig\int_C dx_\mu A_\mu\right)\right\rangle\,.
\end{equation}
 In general $ W(C) $ accumulate divergences (self-energies) that can be eliminated consistently by multiplicative renormalization of the strong coupling $g$ and gauge field $A_\mu$. However, the contour may include cusps where the derivatives of points are not smooth, leading to new cusp divergences~\cite{Polyakov:1980ca,Dotsenko:1979wb,Brandt:1982gz,Korchemskaya:1992je}. These divergences are sensitive to the cusp angle $ \chi$
 (rapidity angle). A standard procedure for the removal of these divergences has been discussed in~\cite{Brandt:1981kf}. These geometrical singularities, along with their corresponding anomalous dimensions, are important in defining the asymptotic states of partonic hard processes \cite{Korchemsky:1993hr}. More specifically, 
for a contour with a single cusp $C_\chi$, the renormalized Wilson loop
reads
\begin{equation}
\begin{aligned}
     &W_R(C_\chi, g_R, \mu)\\
     =& \lim_{\epsilon \to 0} Z^{-1}_{\it cusp}(g_R,\chi,\mu,\epsilon) W(C_\chi, g_R, \mu, \epsilon)
\end{aligned}
\end{equation}
with the renormalized coupling constant $g_R$. The multiplicative renormalizability to any loop integral with a finite number of cusps has been proven in~\cite{Brandt:1981kf,Korchemskaya:1992je}.

\begin{figure*}
    \centering
    \subfloat[\label{wl_DY}]{\includegraphics[width=0.3\linewidth]{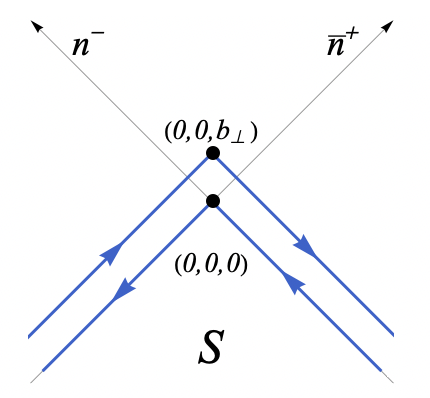}}
    \hfill
    \subfloat[\label{wl_ee}]{\includegraphics[width=0.3\linewidth]{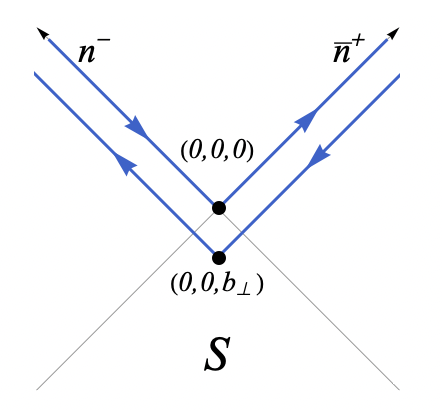}}
     \hfill
    \subfloat[\label{wl_DIS}]{\includegraphics[width=0.3\linewidth]{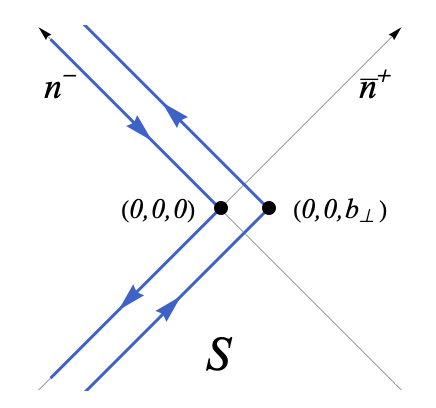}}
    \caption{The Contours of the Wilson lines for the soft function of the cross section in (a) Drell-Yan,  (b) electron-positron annihilation,
    and (c) SIDIS. Hadron $A$ is moving near the light-cone direction $n$,
    and hadron $B$ is moving near the  light-cone direction $\bar{n}$~\cite{Vladimirov:2014hla}.}
    \label{fig:wl}
\end{figure*}

\subsection{Soft functions}
One of the important applications of Wilson loops with cusps is the soft function in the context of TMD factorization. This function 
is composed  of half-infinite Wilson lines, with fixed  initial origin and direction~\cite{Collins:2011zzd}. These lines resum the soft gluon insertions and enforce gauge invariance.

Generically, the half-infinite
Wilson line with the initial point $x$ and  direction $v$ is defined as
\begin{equation}
\begin{aligned}
    W_v(x) =& W[v\infty + x; x ]\\
    =& \mathcal{P} \exp \left[ig \int_0^\infty dzv_\mu A^\mu(vz + x)\right]\,.
\end{aligned}
\end{equation}
The half-infinite Wilson line pointing in the opposite direction is given by
\begin{equation}
\begin{aligned}
    W_{-v}(x) =& W[-v\infty + x; x ]\\
    =& \mathcal{P} \exp \left[ig \int_0^{-\infty} dzv_\mu A^\mu(vz + x)\right]\,,
\end{aligned}
\end{equation}
where $W_v(x)$ denotes the resummation of the eikonalized gluons emitted by the final quark state  moving in the direction $v_\mu$. $W^\dagger_{-v}(x)$ represents the resummation of the eikonalized gluons emitted by the initial quark state moving in the direction $v_\mu$. 

The Wilson lines in the soft functions are process dependent \cite{Collins:2011zzd,Ji:2004wu}. Resummation of the intermediate soft gluons from the initial state hadrons leads to the Wilson lines defined on the past light-cone and final state resummation of the soft gluon gives the Wilson lines defined on the future light-cone.

The Wilson loop defining the soft function for the DY process is illustrated in Fig.~\ref{wl_DY}. The corresponding soft-function for the DY kinematics 
with support on the past light-cone is
\begin{equation}
\begin{aligned}
\label{tmd_soft-1}
&S^{(--)}(b_\perp)=\\
&\frac1{N_c}\Tr\left\langle W_{-\bar n}(0)W^\dagger_{-n}(0)W_{-n}(b_\perp)W^\dagger_{-\bar n}(b_\perp)\right\rangle\,,
\end{aligned}
\end{equation}
where $\bar{n}$ and $n$ denotes the two light-cone directions of the moving hadrons involving in the process, given by the unit vectors $n=(1,0,0_\perp)$ and $\bar n=(0,1,0_\perp)$ in light-cone signature. The time ordering is not needed, since the initial points of the Wilson lines are space-like separated. 

The Wilson loop defining the soft function for the $e^+e^-$ annihilation  process is illustrated in Fig.~\ref{wl_ee}. The soft factor in the annihilation process has support in the forward light cone, with 
\begin{equation}
\begin{aligned}
\label{tmd_soft-3}
    &S^{(++)}(b_\perp)=\\
    &\frac1{N_c}\Tr\langle W_{\bar n}(0)W^\dagger_{n}(0)W_{n}(b_\perp)W^\dagger_{\bar n}(b_\perp)\rangle\,,
\end{aligned}
\end{equation}
which again does not need time ordering. 

The Wilson loop defining the soft function for the SIDIS annihilation  process is illustrated in Fig.~\ref{wl_DIS}. The soft factor is composed of Wilson lines on the past-light-cone and future light-cone,
\begin{equation}
\begin{aligned}
\label{tmd_soft-2}
    &S^{(-+)}(b_\perp)=\\
    &\frac1{N_c}\Tr\left\langle{\cal P}\bigg( W_{-\bar n}(0)W^\dagger_{n}(0)W_{n}(b_\perp)W^\dagger_{-\bar n}(b_\perp)\bigg)\right\rangle
\end{aligned}
\end{equation}
In this case, not all distances are space-like, so the  time-ordering cannot be eliminated.
The Wilson line along $\bar n$ resums the soft gluons emitted from the initial hadron moving in $\bar n$, and the Wilson line along $n$ resums the gluons from the final hadron.
Note that all the soft functions depend only on the transverse coordinate $b_\perp$.

\subsection{Renormalization}
Generally the soft functions defined in \eqref{tmd_soft-1}, \eqref{tmd_soft-3}, and \eqref{tmd_soft-2} have light-cone divergences that can be regulated by tilting $n$ and $\bar n$ slightly off-lightcone~\cite{Collins:2011zzd}. Thus the soft functions form a Wilson loop with two cusp angles on each side, with the rapidity angle defined as $$\cosh\chi=n\cdot\bar n\,.$$
With this in mind, the renormalized soft function can be parameterized as~\cite{Vladimirov:2017ksc}
\begin{equation}
\label{eq:sot_func}
S(\mu,b_\perp,\chi)=\exp\left[K(\mu,b_\perp,\chi)+P(\mu,b_\perp)\right]\,,
\end{equation}
where $K(\mu,b_\perp,\chi)$ is the cusp-angle kernel and $P(\mu,b_\perp)$ is the scheme-dependent non-cusp term that solely depends on $b_\perp$. In a scheme-independent model such as the ILM, we can consider $P(\mu,b_\perp)=0$ for simplicity. Such consideration does not affect the CS kernel extraction since only the $\theta$-dependence is required to determine the CS kernel. 

As we noted earlier, the soft function includes UV divergences that cannot be eliminated by multiplicative renormalization of the strong coupling and gauge field. Around the cusps where the derivatives of points are not smooth, the light-cone divergences further introduce a dependence on the rapidity angle $ \chi $.

With this in mind, the renormalization group (RG) equation of the soft function, a Wilson loop with a Minkowski cusp angle $\chi$, reads
\begin{equation}
    \frac{dK(\mu,b_\perp,\chi)}{d \ln\mu^2}= -\gamma_{K}(\chi,\alpha_s)\,,
\end{equation}
where
\begin{equation}
    \frac{d}{d\ln\mu}=\left(\mu\frac{\partial}{\partial \mu}+\beta(g)\frac{\partial}{\partial g}\right)\,.
\end{equation}
The angle dependence can be factorized out under different limits~\cite{Grozin:2015kna,Korchemsky:1987wg}.
When the angle approaches zero, we have
\begin{equation}
\label{KBX}
    K(\mu,b_\perp,\chi)\xrightarrow{\chi\rightarrow0}K_{B}(\mu,b_\perp)\chi^2\,.
\end{equation}
The corresponding cusp anomalous dimension is called bremsstrahlung function~\cite{Korchemsky:1991zp,Mitev:2015oty}
\begin{equation}
\label{KCSX}
    \gamma_{K}(\chi,\alpha_s) \xrightarrow{\chi\rightarrow0} B(\alpha_s)\chi^2\,.
\end{equation}
On the other hand, when the angle approaches infinity, we have
\begin{equation}
    K(\mu,b_\perp,\chi)\xrightarrow{\chi\rightarrow\infty}K_{\rm CS}(\mu,b_\perp)\chi\,,
\end{equation} 
where $K_{\rm CS}$ is the CS kernel~\cite{Collins:1981uk,Collins:1981va}. It is a nonperturbative function, which represents the interaction of light-quarks in the
QCD vacuum~\cite{Vladimirov:2020umg}.
The rapidity anomalous dimension (RAD), or CS kernel, defines the scaling properties of
transverse momentum dependent distributions and can be extracted from the experimental data.

Finally, the cusp anomalous dimension in the light-cone limit is defined as
\begin{equation}
    \gamma_{K}(\chi,\alpha_s) \xrightarrow{\chi\rightarrow\infty} \Gamma_{\rm cusp}(\alpha_s)\chi\,.
\end{equation}
The perturbative contributions to the light-cone cusp anomalous dimension are given in Appendix~\ref{Appx:cusp}.

\begin{figure}
    \centering
    \includegraphics[width=0.6\linewidth]{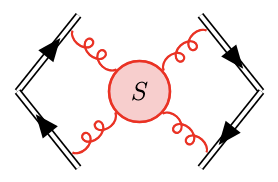}
    \caption{Soft function}
    \label{fig:soft_wilson}
\end{figure}

\section{Soft function in ILM}
\label{SECIII}
For a generic Wilson loop, the nonperturbative contribution  at low resolution (following from say gradient cooling) can be estimated using 
an ensemble of topologically active gauge configurations made chiefly of instantons and anti-instantons. In the dilute regime, the full instanton contribution to the Wilson loop is given by the exponent of the all-order single instanton result~\cite{Shuryak:2000df,Shuryak:2021hng,Dorokhov:2002qf}
\begin{equation}
\label{loop_inst}
    W(C)=\exp\left[\frac{1}{N_cV}\sum_I\int d^4z ~\mathrm{Tr}_c\left(W_I(\rho,z)-1\right)\right]
\end{equation}
with the single instanton insertion
\begin{equation}
W_I(\rho,z)=e^{i\tau^a\phi_I^a(\rho,z)}\,,
\end{equation}
where the single-instanton profile in the soft Wilson loop is defined as 
\begin{equation}
\label{single_inst_wilson}
        \phi_I^a(\rho,z)=\int_{C} dx_\mu  \frac{\bar\eta^a_{\mu\nu}(x-z)_\nu\rho^2}{(x-z)^2[(x-z)^2+\rho^2]}\,.
\end{equation}
$\eta^a_{\mu\nu}$ is the 't Hooft symbol defined as \cite{Liu:2024rdm,Vainshtein:1981wh}
\begin{equation}
    \bar\eta^{a}_{\mu\nu}=\begin{cases}
        \epsilon^{a}{}_{\mu\nu} ,\ & \mu\neq4,\ \nu\neq4\\
        -\delta^{a}_{\mu} ,\ & \mu\neq4,\ \nu=4 \\
        \delta^{a}_\nu ,\ & \mu=4,\ \nu\neq4
    \end{cases}
\end{equation}
and its  conjugate $\eta^{a}_{\mu\nu}=\bar{\eta}^{a}_{\mu\nu}$ and $\eta^{a}_{\mu4}=-\bar{\eta}^{a}_{\mu4}$ for $\mu\neq4,\ \nu\neq4$. The definition for anti-instantons can be obtained by $\bar\eta^a_{\mu\nu}\leftrightarrow\eta^a_{\mu\nu}$. The exponentiation
arises due to taking into account the many-instanton configurations effect.
Note that the path ordering operator $\mathcal{P}$ in \eqref{loop} is not needed, since the instanton field is a hedgehog in
color space, with the color-space orientations locked.

\begin{figure}
    \centering
\subfloat[]{\includegraphics[width=0.5\linewidth]{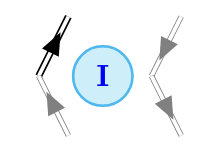}}
\hfill
\subfloat[]{\includegraphics[width=0.5\linewidth]{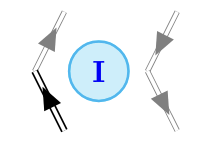}}
\hfill
\subfloat[]{\includegraphics[width=0.5\linewidth]{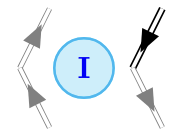}}
\hfill
\subfloat[]{\includegraphics[width=0.5\linewidth]{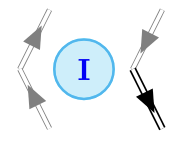}}
    \caption{(a) path $L_1$ is labeled by $v_1$ and $b_1$; (b)  path $L_2$ is labeled by $v_2$ and $b_2$; (c) path $L_3$ is labeled by $v_3$ and $b_3$; (d) path $L_4$ is labeled by $v_4$ and $b_4$}.
    \label{fig:wilson_lines}
\end{figure}

In the ILM, the soft function associated to Fig.~\ref{fig:soft_wilson} 
is composed of four contributions,
\begin{widetext}
\begin{equation}
\begin{aligned}
\label{soft_wilson}
        \phi^a(z,b_\perp,\rho)=&\bar{\eta}^a_{\mu\nu}v_\mu z_\nu \int_0^\infty ds \varphi_v(s,z,\rho)+\bar{\eta}^a_{\mu\nu}\bar{v}_\mu z_\nu\int_{-\infty}^0 ds \varphi_{\bar v}(s,z,\rho)\\
        &-\bar{\eta}^a_{\mu\nu}v_\mu (z-b_\perp)_\nu \int_0^\infty ds \varphi_v(s,z-b_\perp,\rho)-\bar{\eta}^a_{\mu\nu}\bar{v}_\mu (z-b_\perp)_\nu \int_{-\infty}^0 ds \varphi_{\bar v}(s,z-b_\perp,\rho)\,,
\end{aligned}
\end{equation}
where the instanton profiling function along the Wilson line is defined as
\begin{equation}
\begin{aligned}
    \varphi_{v}(s,z,\rho)=\frac{\rho^2}{(s^2-2v\cdot zs+z^2)(s^2-2v\cdot zs+z^2+\rho^2)}\,.
\end{aligned}
\end{equation}
\end{widetext}
As we noted for the generic Wilson loop \eqref{loop_inst}, there is no need
for the path ordering. 
Each term in \eqref{soft_wilson} represents a half-infinite segment in the Wilson loop in Fig.~\ref{fig:wilson_lines}.
In terms of (\ref{soft_wilson}), the  Wilson loop related to the soft function can be written as
\begin{equation}
\begin{aligned}\
\label{eq:W}
    &W(\rho,b_\perp,\theta)=\\
    &\exp\left[\frac{2n_{I+A}}{N_c}\int d^4z \left(\cos\phi(z,\rho,b_\perp,\theta)-1\right)\right]\,,
\end{aligned}
\end{equation}
where the Euclidean angle is identified as $v\cdot \bar{v}=\cos\theta$, which can be analytically continued to the Minkowskian rapidity through
$\theta\rightarrow i\chi$. 
The norm  $\phi=\sqrt{(\phi^a)^2}$ follows from~\eqref{soft_wilson}. 

When $\theta=0$, only the Wilson line self-energy and the Coulomb-type interaction between the parallel Wilson lines contribute. Following translational symmetry in  Euclidean space, this integral will reduce from the four dimensional $z$-integral  to three dimensional integral (one longitudinal and two transverse directions) with linear dependence on $z_4$,
\begin{equation}
\begin{aligned}
\ln W(\rho,b_\perp,0)=&-\int dz_4 V_{\rm Coul}(b_\perp/\rho)\,,
\end{aligned}
\end{equation}
where the Coulomb potential is defined as
\begin{widetext}
\begin{equation}
\label{V_inst}
V_{\rm Coul}(b_\perp/\rho)=\frac{2n_{I+A}}{N_c}\int dz_3d^2z_\perp \left[1-\cos\phi(z,\rho,b_\perp,0)\right]\,.
\end{equation}
\end{widetext}
This linear $z_4$ divergence reflects on the self-energy in the Wilson line. After subtracting the linear divergent and angle-independent term from the exponent in \eqref{eq:W}, the remaining contribution comes from the cusp-angle,
\begin{widetext}
\begin{equation}
\begin{aligned}
\label{K}
K(\rho,b_\perp,\theta)=\ln W(\rho,b_\perp,\theta)-\ln W(\rho,b_\perp,0)=&\frac{2n_{I+A}}{N_c}\int d^4z \left[\cos\phi(z,\rho,b_\perp,\theta)-\cos\phi(z,\rho,b_\perp,0)\right]\,.
\end{aligned}
\end{equation}
\end{widetext}
The subtraction ensures $K=0$ for $\theta\rightarrow0$. The soft function can be defined by the exponentiated kernel $K$
\begin{equation}
S(b_\perp,\rho,\theta)\equiv\frac{W(b_\perp/\rho,\theta)}{W(b_\perp/\rho,0)}=e^{K(\rho,b_\perp,\theta)}
\end{equation}
with $P=0$ as defined in~\eqref{eq:sot_func}.


For simplicity, we set the zero angle velocity along the $z_4$-axis, i.e. $v\cdot z|_{\theta=0}=\bar v\cdot z|_{\theta=0}=z_4$.  The instanton squared profile function at zero angle reads
\begin{widetext}
\begin{equation}
\begin{aligned}
\phi^2(z,\rho,b_\perp,\theta=0)=&-2\pi^2\frac{z_3^2+z_\perp^2-\frac14b^2_\perp}{\sqrt{z_3^2+\left(z_\perp+\frac12b_\perp\right)^2}\sqrt{z_3^2+\left(z-\frac12b_\perp\right)^2}}\\
        &\times\left(1-\frac{\sqrt{z_3^2+\left(z_\perp+\frac12b_\perp\right)^2}}{\sqrt{z_3^2+\left(z_\perp+\frac12b_\perp\right)^2+\rho^2}}\right)\left(1-\frac{\sqrt{z_3^2+\left(z_\perp-\frac12b_\perp\right)^2}}{\sqrt{z_3^2+\left(z_\perp-\frac12b_\perp\right)^2+\rho^2}}\right)\\
        &+\pi^2\left(1-\frac{\sqrt{z_3^2+(z_\perp+\frac12b_\perp)^2}}{\sqrt{z^2_3+(z_\perp+\frac12b_\perp)^2+\rho^2}}\right)^2+\pi^2\left(1-\frac{\sqrt{z_3^2+(z_\perp-\frac12b_\perp)^2}}{\sqrt{z^2_3+(z_\perp-\frac12b_\perp)^2+\rho^2}}\right)^2\,.
\end{aligned}
\end{equation}
\end{widetext}
The UV divergence as $z\rightarrow0$ is regulated by the instanton size $\rho$, and the IR divergence $z\rightarrow\infty$ cancels out between the two-anti-parallel Wilson lines. The absence of the $z_4$ dependence is due to rotational symmetry in Euclidean space.

To evaluate the cusp-angle dependence, we adopt the symmetric parameterization for the angle
\begin{align}
v_\mu=&\left(\cos\frac{\theta}2,\sin\frac\theta2,0_\perp\right) \,,\nonumber\\
\bar v_\mu=&\left(\cos\frac{\theta}2,-\sin\frac\theta2,0_\perp\right)\,,
\end{align}
with the squared norm profile
\begin{widetext}
\begin{equation}
\begin{aligned}
\phi^2(z,\rho,b_\perp,\theta)=&\sum_{i,j=1}^4\frac{v_i\cdot v_j(z-b_i)\cdot(z-b_j)- (v_i\cdot z)(v_j\cdot z)}{{\sqrt{(z-b_i)^2-(v_i\cdot z)^2}\sqrt{(z-b_j)^2-(v_j\cdot z)^2}}}w_i\left(z-b_i,\rho,v_i\right)w_j\left(z-b_j,\rho,v_j\right)\,.
\end{aligned}
\end{equation}
The sum runs over the the four paths $L_i$ in Fig.~\ref{fig:wilson_lines}, with the symmetric parameterization
\begin{align}
    b_1=b_2=&-\frac12b_\perp\,, & b_3=b_4=&\frac12b_\perp\,,\\
    v_1=v_3=&v\,, & v_2=v_4=&\bar v\,,
\end{align}
and for each path $L_i$ the instanton profiles are defined as
\begin{align}
    w_1(z,\rho,v)=&w(z,\rho,v)\,, & w_2(z,\rho,v)=&w(z,\rho,-v)\,, \nonumber\\
    w_3(z,\rho,v)=&-w(z,\rho,v)\,, & w_4(z,\rho,v)=&-w(z,\rho,-v) \,,
\end{align}
with
\begin{equation}
\begin{aligned}
    w(z,\rho,v)=&\frac{\pi}{2}\left(1-\frac{\sqrt{z^2-(v\cdot z)^2}}{\sqrt{z^2+\rho^2-(v\cdot z)^2}}\right)+\tan^{-1}\left(\frac {v\cdot z}{\sqrt{z^2-(v\cdot z)^2}}\right)\\
    &
    -\frac{\sqrt{z^2-(v\cdot z)^2}}{\sqrt{z^2+\rho^2-(v\cdot z)^2}}\tan^{-1}\left(\frac {v\cdot z}{\sqrt{z^2+\rho^2-(v\cdot z)^2}}\right)\,.
\end{aligned}
\end{equation}
\end{widetext}

\begin{figure*}
    \centering
\subfloat[\label{fig:COUL_a}]{\includegraphics[width=.5\linewidth]{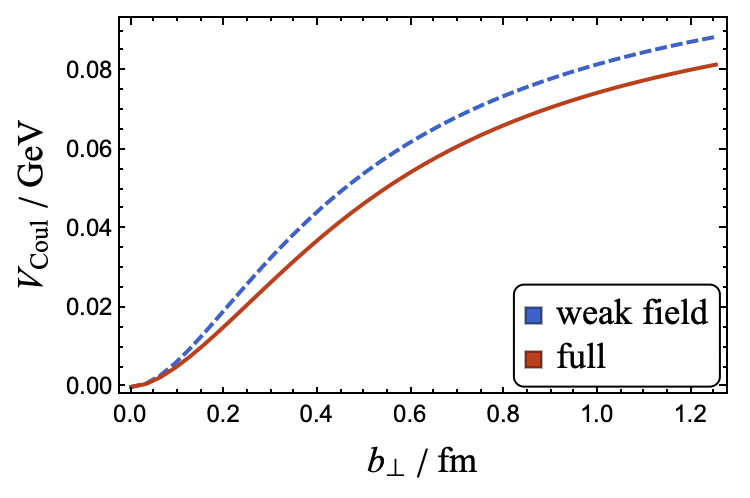}}
\hfill
\subfloat[\label{fig:COUL_b}]{\includegraphics[width=.475\linewidth]{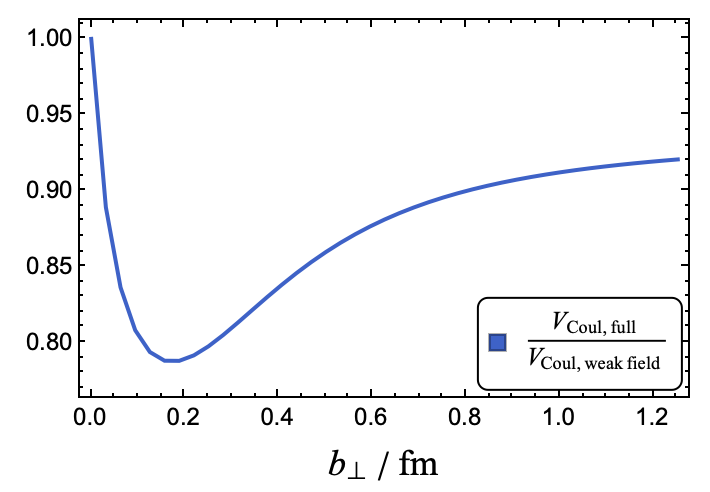}}
    \caption{(a) Instanton liquid estimation of the Coulomb potential between two parallel Wilson lines, with the full calculation \eqref{V_inst} and the weak field approximation \eqref{V_weak}. We also plot their ratio in (b).}
    \label{fig:COUL}
\end{figure*}

\begin{figure*}
    \centering
\subfloat[\label{fig:KCSX_a}]{\includegraphics[width=.5\linewidth]{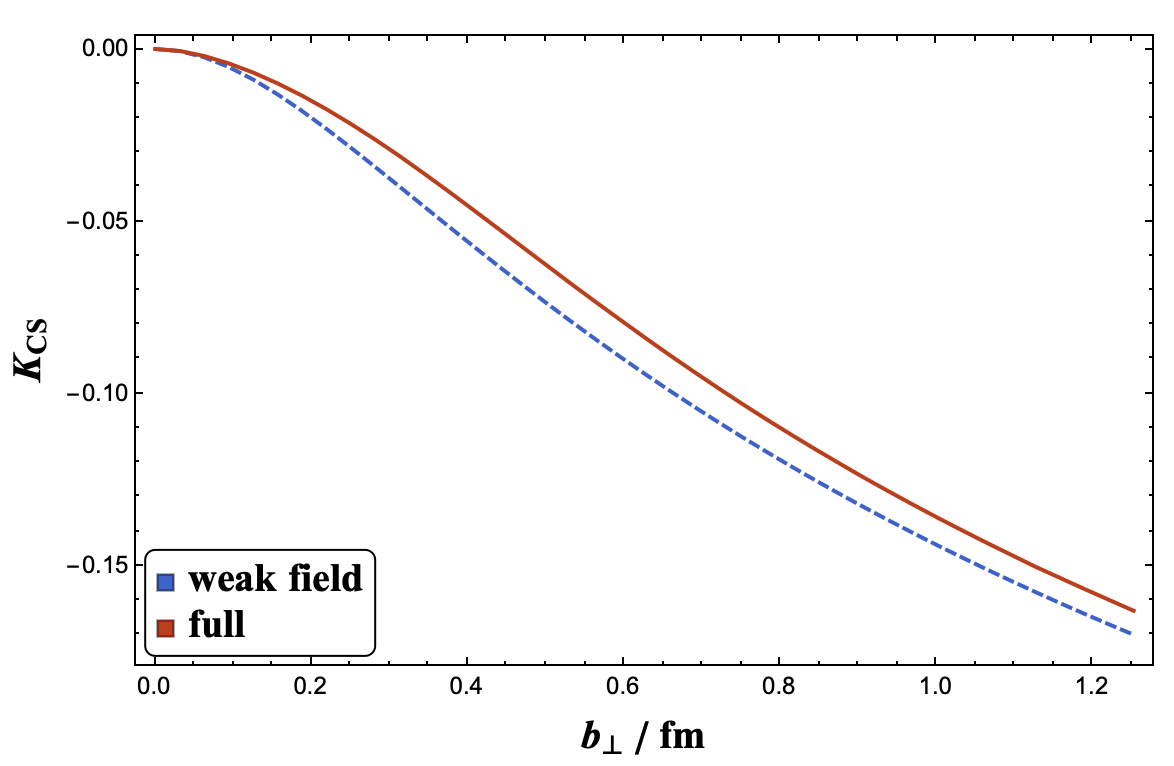}}
\hfill
\subfloat[\label{fig:KCSX_b}]{\includegraphics[width=.46\linewidth]{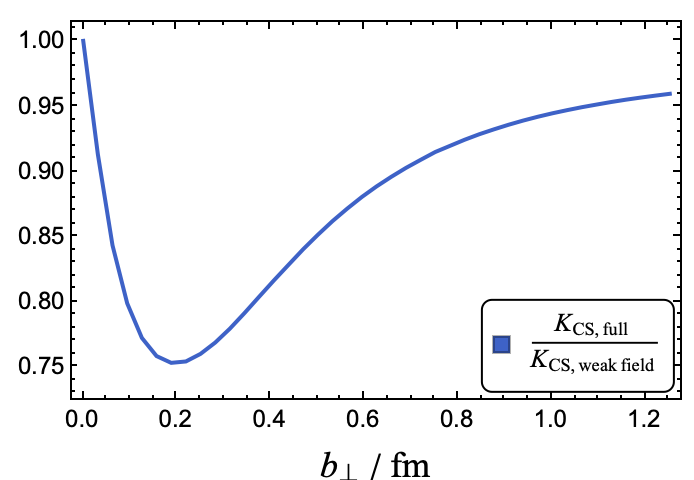}}
    \caption{(a) Instanton liquid estimation on CS kernel with the full calculation \eqref{K} and weak field approximation \eqref{CS_weak}. We also plot their ratio in (b).}
    \label{fig:KCSX}
\end{figure*}

\begin{figure*}
    \centering
\subfloat[\label{fig:theta_a}]{\includegraphics[width=.5
\linewidth]{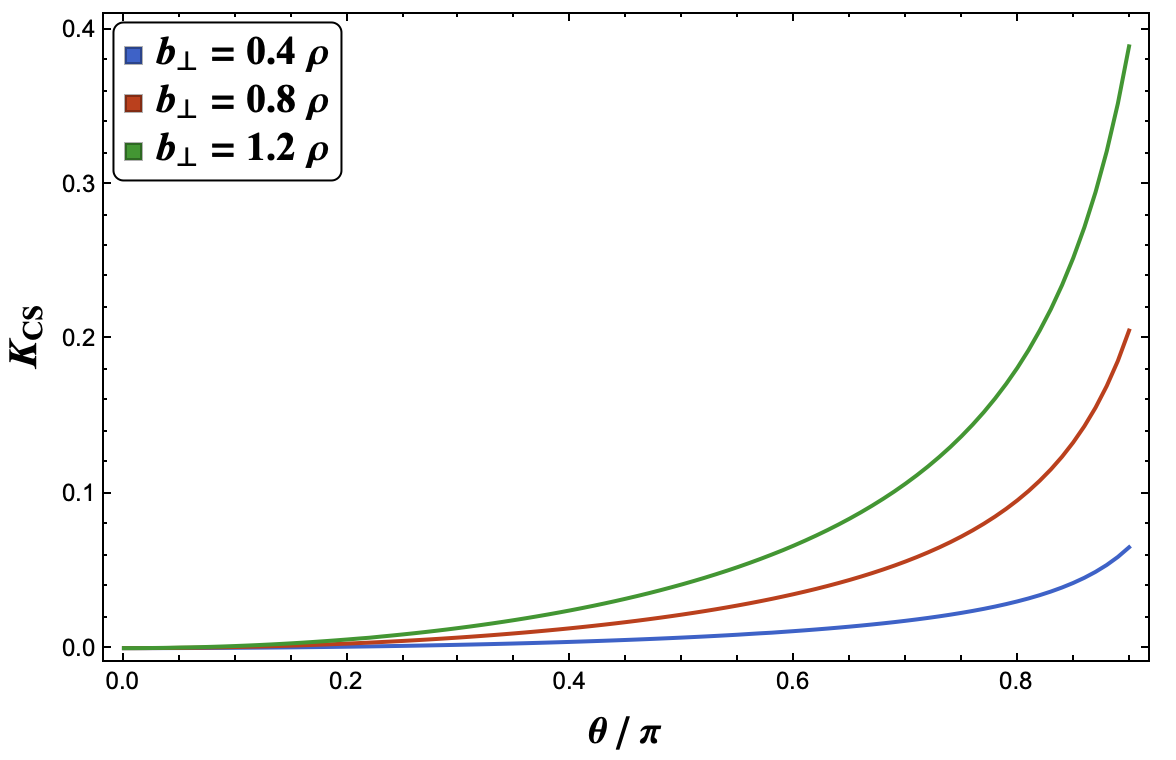}}
\hfill
\subfloat[\label{fig:theta_b}]{\includegraphics[width=.5\linewidth]{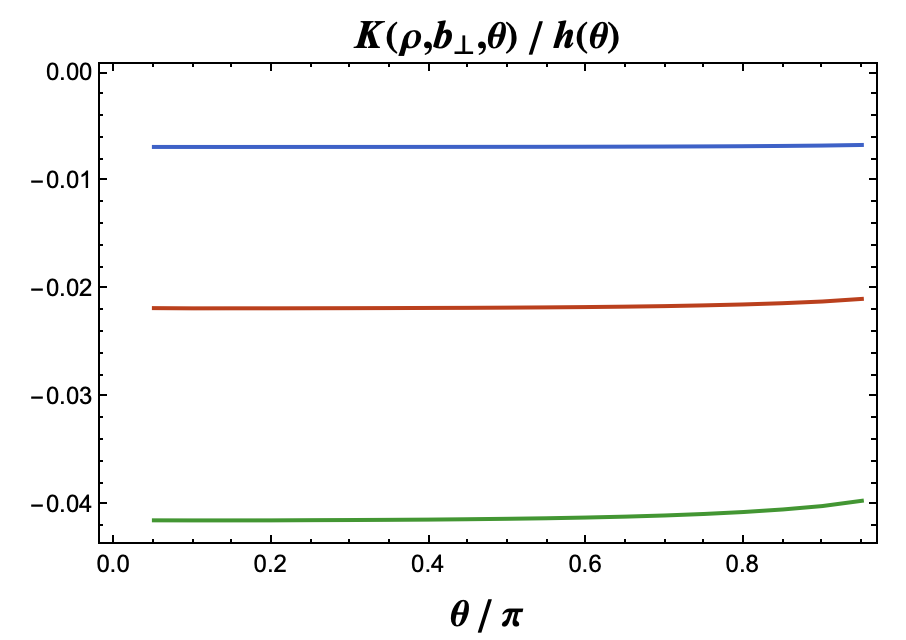}}
    \caption{The angular dependence of $K$ in \eqref{K}. }
    \label{fig:theta}
\end{figure*}

\subsection{Weak field}
A full analytical analysis of the results is too involved and beyond the scope of this work. Below, we will rely instead on a numerical evaluation. However, in the weak field limit, the analytical form can be worked out for an initial estimate. Indeed,  in the weak field limit ($|\phi|\ll1$), we have
\begin{widetext}
\begin{equation}
    \ln W(\rho,b_\perp,\theta)\simeq\ln W^{(1)}(\rho,b_\perp,\theta)\equiv-\frac{n_{I+A}}{N_c}\int d^4z \phi^2(z,\rho,b_\perp,\theta)\,,
\end{equation}
where the integral can be rewritten as
\begin{equation}
\begin{aligned}
\label{SUMLI}
    \int d^4z\phi^2=&\sum_{i,j=1}^4\int_{L_i} ds\int_{L_j}ds'\bigg\{v_i\cdot v_jd_1[(sv_i-s'v_j)^2+b^2_\perp]\\
    &-\frac{ss'(1-(v_i\cdot v_j)^2)+v_i\cdot v_j b_\perp^2}{s^2+s'^2-2v_i\cdot v_jss'+b_\perp^2}d_2[(sv_i-s'v_j)^2+b^2_\perp]\bigg\}\,.
\end{aligned}
\end{equation}
\end{widetext}
with the instanton profiling functions $d_1$ and $d_2$  defined as
\begin{equation}
    d_1(x^2)=\frac{3\pi^2\rho^4}x\int_0^\infty dk\frac{J_2(kx)}{kx}\mathcal{F}^2_g(\rho k)\,,
\end{equation}
\begin{equation}
    d_2(x^2)=\frac{\pi^2\rho^4}x\int_0^\infty dkJ_3(kx)\mathcal{F}^2_g(\rho k)\,,
\end{equation}
and 
\begin{equation}
    \mathcal{F}_g(k)=\frac4{k^2}-2K_2(k)\,.
\end{equation}
The sum in (\ref{SUMLI}) runs over the the four paths $L_i$ shown in Fig.~\ref{fig:wilson_lines}, using the symmetric parameterization
\begin{align}
    v_1=v_3=&v\,, & v_2=v_4=&\bar v\,.
\end{align}
For each path $L_i$, the $s$-integral is defined as
\begin{align}
    \int_{L_1}ds=&\int_0^\infty ds\,, & \int_{L_2}ds=&\int_{-\infty}^0 ds\,, \nonumber\\
    \int_{L_3}ds=&\int_{\infty}^0 ds\,,  & \int_{L_4}ds=&\int^{-\infty}_0 ds\,.
\end{align}
The details of this integral are given in Appendix~\ref{soft_wilson_int}.
The $s$-integral allows us to write down the analytical form of the Coulomb potential and the CS kernel in the weak field limit.  

In the zero angle limit, the two parallel Wilson lines give rise to the Coulomb potential,
\begin{equation}
\begin{aligned}
\label{V_weak}
    &V^{(1)}_{\rm Coul}=\frac{4n_{I+A}\pi^2\rho^4}{N_c}\\
    &\quad \times \int_0^\infty dk\mathcal{F}^2_g(\rho k)\left(1-\frac{\sin kb_\perp}{kb_\perp}\right)\,,
\end{aligned}
\end{equation}
with the UV divergence cutoff by the instanton size $\rho$. 
In Fig.~\ref{fig:COUL_a} we show the result of the  Coulomb self-energy 
for the leading order (LO) weak field limit \eqref{V_weak} 
dashed-blue line, the next-to-leading order (NLO)
weak field limit dashed-brown line, and the full and strong field limit 
\eqref{V_inst} solid-green line. The
NLO result is in good agreement with the full result over a large range of separations $b_\perp$. 
In Fig.~\ref{fig:COUL_b} the ratio of the LO to full Coulomb self-energy is shown versus 
$b_\perp$, where the deviation is mostly at small separations. The accumulated self-energy at large separation is about $80\,\rm MeV$ in agreement with earlier estimates~\cite{Shuryak:2021fsu}.

To evaluate the contribution to the CS kernel in the weak field limit, it is useful to  change variables $\sigma=s-s'$ and $\eta=s/\sigma$ to separate the angle dependence from the rest of the integral, 
\begin{equation}
\label{csX}
    K^{(1)}(\rho,b_\perp,\theta)=K_{\rm CS}^{(1)}(b_\perp/\rho)h(\theta)\,,
\end{equation}
with the cusp factor
$$h(\theta)=\theta \cot\theta-1$$
and the weak field kernel
\begin{equation}
\begin{aligned}
\label{CS_weak}
    &K_{\rm CS}^{(1)}=\frac{4n_{I+A}\pi^2\rho^4}{N_c} \\
    &\qquad \times \int_0^\infty \frac{dk}k\mathcal{F}^2_g(\rho k) \left(J_0(kb_\perp)-1\right)\,.
\end{aligned}
\end{equation}
This separation is exact in the weak field limit, as we detail in
Appendix~\ref{soft_wilson_int}. In the large $b_\perp$ limit, the asymptotic form of \eqref{CS_weak} in weak field reduces to a logarithmic function in $b_\perp$,
\begin{equation}
    K^{(1)}_{\rm CS}\xrightarrow{b_\perp\rightarrow\infty} -\frac{4n_{I+A}\pi^2\rho^4}{N_c}\left[\ln (b_\perp/\rho) +1\right]\,.
\end{equation}

\subsection{Strong field}
In the strong field limit, the evaluation of the soft function (\ref{eq:W}) can only be carried out numerically. In Fig.~\ref{fig:KCSX_a} we show the ILM results for the CS kernel at low resolution, from
the full or strong field limit \eqref{K} red-solid curve, in comparison to the weak field limit \eqref{CS_weak}, versus the separation $b_\perp$. The same comparison for the ratio is shown in 
Fig.~\ref{fig:KCSX_b}. The deviation between the full and the LO result is mostly at small separations,
where the field is strong.

The numerical dependence on the cusp factor for the full result follows closely  $h(\theta)$, as illustrated in~Fig.~\ref{fig:theta_b}.  We  conclude that the weak field separation (\ref{csX}) holds in the strong field limit as well, 
\begin{equation}
\label{K}
K(\rho,b_\perp,\theta)\simeq K_{\rm CS}(b_\perp/\rho)h(\theta)\,.
\end{equation}
In  Fig.~\ref{fig:theta_a} we show the dependence of the numerical kernel
on $b_\perp$ in units of the instanton size $\rho$, for varying angle $\theta/\pi$. As expected, the dependence drops out at small angles. For future reference, we list the numerical values for $K_{\rm CS}$ \eqref{K} in Table \ref{tab:CS_numbers}.

\begin{table}
\centering
\begin{minipage}{0.44\linewidth}
\begin{tabular}{@{}cc@{}}
\toprule
`$b_\perp$ (GeV$^{-1}$) & $K_{\rm CS}$ \\ \midrule
0.000  &  0.000 \\
0.174  & -0.006 \\
0.348  & -0.022 \\
0.522  & -0.045 \\
0.695  & -0.074 \\
0.869  & -0.108 \\
1.043  & -0.146 \\
1.217  & -0.188 \\
1.565  & -0.284 \\
1.912  & -0.390 \\
2.260  & -0.503 \\
2.608  & -0.618 \\
2.955  & -0.733 \\
3.303  & -0.846 \\
3.651  & -0.954 \\
3.998  & -1.058 \\
4.346  & -1.157 \\ \bottomrule
\end{tabular}
\end{minipage} 
\hfill
\begin{minipage}{0.44\linewidth}
\centering
\begin{tabular}{@{}cc@{}}
\toprule
$b_\perp$ (GeV$^{-1}$) & $K_{\rm CS}$ \\ \midrule
4.694  & -1.251 \\
5.042  & -1.340 \\
5.389  & -1.424 \\
5.737  & -1.504 \\
6.085  & -1.580 \\
6.432  & -1.653 \\
6.780  & -1.722 \\
7.128  & -1.787 \\
7.475  & -1.850 \\
7.823  & -1.910 \\
8.171  & -1.969 \\
8.519  & -2.025 \\
8.866  & -2.078 \\
9.214  & -2.129 \\
9.562  & -2.179 \\
9.909  & -2.226 \\
10.257 & -2.273 \\ \bottomrule
\end{tabular}
\end{minipage}
\caption{Table of the CS kernel $K_{\rm CS}(b_\perp)$ as a function of $b_\perp$.}
\label{tab:CS_numbers}
\end{table}

The ILM result (\ref{K}) can be analytically continued to Minkowski signature
 $\theta\rightarrow i\chi$, 
\begin{equation}
    v\cdot \bar v=\cos\theta\rightarrow\cosh\chi\simeq1+\frac{Q^2}{2m_h^2}\,,
\end{equation}
with the universal cusp factor
$h(\chi)\simeq\ln Q^2/m_h^2$.
Here $m_h$ is the hadron mass, and $Q$ is the momentum transfer in the hadron, assuming  $v_\mu=p_\mu /m_h$ for simplicity. 
A comparison of (\ref{K}) to (\ref{KBX}) shows that in the ILM
\begin{equation}
    K_B\simeq-\frac13 K_{\rm CS}\,.
\end{equation}

The factorization of angular dependence in \eqref{K} over the full range of $\theta$ ensures that analytic continuation of $\theta$ does not alter the $b_\perp$-dependence of the CS kernel. 
This property is generally absent in perturbative calculations, leading to different cusp anomalous dimension beyond one-loop in Euclidean and Minkowskian spaces \cite{Ji:2019sxk}. Our prediction suggests that in lattice QCD, the Euclidean soft function at large \(b_\perp\) can be computed with real $\theta$, and the CS kernel can be extracted by factoring out $h(\theta)$.

Furthermore, understanding the asymptotic form at large $b_\perp$ is of importance for the phenomenological extractions, as various groups employ different models for $b_\perp$. From Fig.~\ref{fig:KCSX_b} we observe that the CS kernel agrees with the weak field limit both for $b_\perp/\rho\ll 1$ (kinematics) and $b_\perp/\rho\gg 1$ (asymptotics). The deviation from the weak field limit is only about $10\%$ for $b_\perp/\rho\sim 1$.
The large-$b_\perp$ asymptotic behavior of the CS kernel is found to be logarithmic, in agreement with an early phenomenological paramterization used in~\cite{Collins:1985xx}. Thus, our ILM calculations provide strong constraints on the models for phenomenological fitting.

\begin{figure}
  \centering
    \includegraphics[width=\linewidth]{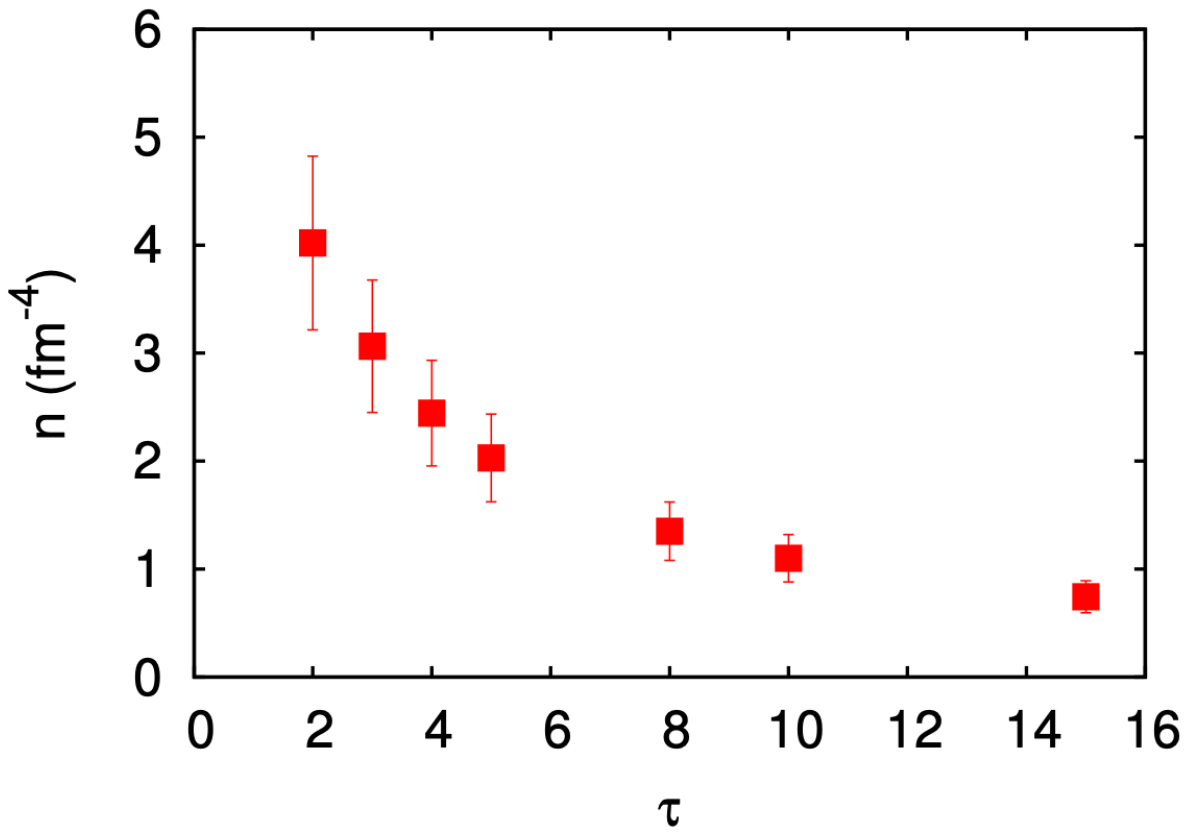}
    \caption{Instanton density $n$ as a function of the dimensionless cooling time $\tau$~\cite{Athenodorou:2018jwu} where $\tau=t/a^2$ with the lattice space $a=0.139$ fm.}
    \label{fig:COOL}
\end{figure}

\begin{figure*}
    \centering
\subfloat[\label{CS1}]{\includegraphics[width=0.5\linewidth]{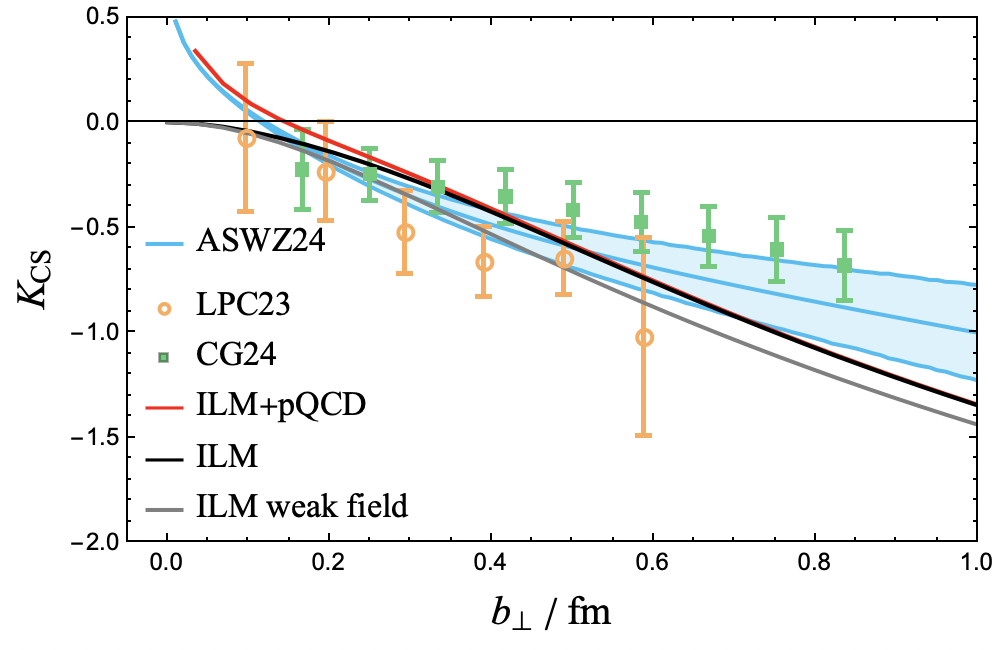}}
\hfill
\subfloat[\label{CS2}]{\includegraphics[width=0.5\linewidth]{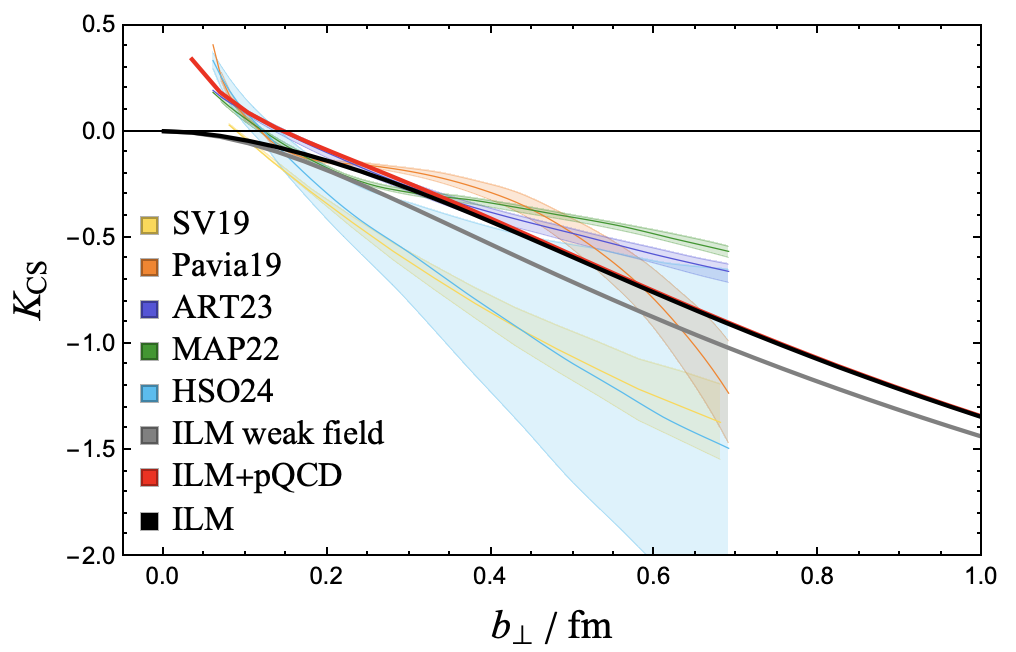}}
    \caption{(a) The black-solid curve is the  ILM result for the CS kernel with $\rho=0.343$ fm, $n_{I+A}=7.46$ fm$^{-4}$, compared to  the weak field approximation gray-soli curve and the optimized combination (\ref{KINTER}-\ref{b_star}) with the short-distance perturbative contribution red-solid curve . The results are compared to the recent lattice calculation by ASWZ24~\cite{Avkhadiev:2023poz},  a new lattice calculation using Coulomb-gauge TMD correlators~\cite{Zhao:2023ptv} by CG24~\cite{Bollweg:2024zet} and results from Lattice Parton Collaboration LPC23~\cite{LatticePartonLPC:2023pdv}. (b) shows the comparison of our results with the phenomenologically  extracted CS kernel  by SV19~\cite{Scimemi:2019cmh}, Pavia19~\cite{Bacchetta:2019sam}, MAP22~\cite{Bacchetta:2022awv}, ART23~\cite{Moos:2023yfa}, and HSO24~\cite{Gonzalez-Hernandez:2022ifv}.}
    \label{fig:CS}
\end{figure*}

\section{Collins-Soper Evolution}
\label{SECIV}
In the perturbative calculation, the Wilson lines in TMD functions introduces rapidity divergence which requires different regularization other than the UV divergences~\cite{Collins:1981uk,Collins:2011zzd,Becher:2010tm,Echevarria:2011epo,Chiu:2012ir,Collins:2008ht,Collins:1992tv,Chiu:2011qc,Manohar:2006nz,Li:2016axz,Ebert:2018gsn}. As a regularization, the path of Wilson lines can be taken off the lightcone~\cite{Collins:2011zzd}, specified by the variable $\zeta$ related to finite rapidity. 
The variable $\zeta$
is the energy of the hadron \cite{Boer:2014tka,Idilbi:2004vb}, 
\bea
    \zeta^2=&&m_h^2x^2e^{2(y-y_v)}\nonumber\\
    =&&\frac{(2xp\cdot v)^2}{v^2}=2(xp^+)^2\frac{v^-}{v^+}\,,
\eea
where $y=\frac12\ln \frac{p^+}{p^-}$ is the hadron rapidity with mass $m_h$, and $y_v=\frac12\ln \frac{v^+}{v^-}$  the off lightcone rapidity velocity $v$. In general, $v$ can be different from the momenta of both involving hadrons $h$ and $h'$ in the TMD scattering process. For the simplest case, we can assume $v_\mu=p_\mu/m_h$ ($y_v=y$), $v_\mu=p'_\mu/m_{h'}$ ($y_v=y'$), or  which simply assume $\zeta^2=Q^2$.

The variation in $\zeta$ is determined by the CS kernel $K_{\rm CS}$ \cite{Collins:1981uk,Collins:1981va}. The integrability of the TMD renormalization group (RG) equations guarantees that
\begin{equation}
\label{RG_K}
    \frac{dK_{\rm CS}(b_\perp,\mu) }{d\ln \mu^2}=-\Gamma_{\mathrm{cusp}}(\alpha_s(\mu))\,.
\end{equation}

\subsection{The Collins-Soper kernel}
The ILM emerges from the QCD vacuum by gradient flow cooling~\cite{Leinweber:1999cw,Athenodorou:2018jwu}. 
In Fig.~\ref{fig:COOL} we show the dependence of the 
instanton density $n$ on the cooling time $t$ from
the lattice analysis in~\cite{Athenodorou:2018jwu}.
The cooling time $t$ can be identified
with the renormalization scale 
$$\mu\sim\frac{1}{\sqrt{8t}}\,.$$
where the cooling time $t$ is defined in terms of the lattice spacing, $\tau=t/a^2$. The instanton density asymptotes $n\sim 1/\rm fm^4$
at large cooling times ($\tau=9$) 
or low resolution $\mu=520~\mathrm{MeV}\sim 1/\rho$, as most instanton molecules are 
resolved. This is the realm where the spontaneous breaking of chiral symmetry is the lore.
At shorter cooling times ($\tau=0.6$) or high resolution $\mu=2~\mathrm{GeV}$, more 
instanton molecules are present, hence the larger density $n\sim 7.46/\rm fm^4$.

With this in mind, we evaluate \eqref{K} using the larger instanton density 
to account for the non-perturbative part of the CS kernel at high resolution and  large separation $b_\perp$. This is in line with the
'dense' ILM suggested in~\cite{Shuryak:2021fsu}.

At small separations $b_\perp\ll b_{\rm max}$, most of the instanton annihilates with anti-instanton, and perturbative effects are dominant, hence the interpolating solution for the evolution equation \eqref{RG_K} is \cite{Collins:2011zzd} 
\begin{equation}
\begin{aligned}
\label{KINTER}
K_{\rm CS}&=K^{(\rm np)}_{\rm CS}(b_\perp,\mu) + K^{(\rm pert)}_{\rm CS}(b_*, \mu_b) \\
&\qquad -2\int_{\mu_b}^\mu \frac{d\mu'}{\mu'}\Gamma_{\rm cusp}(\alpha_s(\mu'))\,.
\end{aligned}
\end{equation}
The non-perturbative part $K^{(\rm np)}_{\rm CS}$ is fixed from the ILM evolved to higher resolution,
\begin{equation}
    K^{(\rm np)}_{\rm CS}(b_\perp,\mu)=K^{(\rm inst)}_{\rm CS}(b_\perp/\rho,\mu) - K^{(\rm inst)}_{\rm CS}(b_*/\rho,\mu)\,.
\end{equation}
The perturbative contribution  $K^{(\rm pert)}_{\rm CS}(b_*, \mu_b)$ (integration constant obtained by solving the RG equation of $K_{\rm CS}$) and cusp anomalous dimension $\Gamma_{\rm cusp}$are given in Appendix \ref{Appx:cusp}. The transverse scale is parameterized by
\begin{equation}
    \mu_b=\frac{2e^{-\gamma_E}}{b^*(b_\perp)}\xrightarrow{b_\perp\rightarrow\infty}\frac{2e^{-\gamma_E}}{b_{\rm max}}
\end{equation}
with the asymptotic saturation following from the conventional $b^*$-parameterization~\cite{Collins:2011zzd},
\begin{equation}
\label{b_star}
    b^*(b_\perp)=\frac{b_\perp}{\sqrt{1+b_\perp^2/b_{\rm max}^2}}\,,
\end{equation}
which reduces to $b_\perp$ in the limit $b_\perp\to0$.
Here $\gamma_E$ is  Euler constant, and the optimal non-perturbative distance is  $b_{\rm max}=0.56$ fm, which is chosen to optimize the interpolation between non-perturbative and perturbative contributions in the lattice results of ASWZ24~\cite{Avkhadiev:2023poz}. 
The lattice results are extracted from the quasi-TMD and matched to the $\overline{\rm MS}$ scheme. Thus our result of the CS kernel depends on the model parameterization \eqref{b_star} and renormalization scheme. Since $b_\perp$ near zero is not available in lattice, the parameterization in \eqref{b_star} is still applicable.

In Fig.~\ref{fig:CS} we show our results for the CS kernel.
The solid-black curve is the full non-perturbative contribution to
the CS from the ILM for $\rho=0.343$ fm and  $n_{I+A}=7.46$ fm$^{-4}$,
compared to the gray-solid curve in the weak-field limit, as a function of $b_\perp$. The deviation between the strong and weak field limits is small. 
The amended result for small $b_\perp$ using the interpolating solution 
(\ref{KINTER}-\ref{b_star})  for the CS kernel with $b_{\rm max}$ parameter, is shown as the solid-red curve. In Fig.~\ref{CS1} our results are compared 
to the recent lattice calculation by ASWZ24~\cite{Avkhadiev:2023poz}, a new lattice calculation using Coulomb-gauge TMD correlators~\cite{Zhao:2023ptv} by CG24~\cite{Bollweg:2024zet} and the results from the Lattice Parton Collaboration LPC23~\cite{LatticePartonLPC:2023pdv}. The ILM are in overall agreement with the lattice results, especially with those reported 
by ASWZ24~\cite{Avkhadiev:2023poz}. In Fig.~\ref{CS2}
 our results are compared with the  phenomenologically  extracted CS kernel by SV19~\cite{Scimemi:2019cmh}, Pavia19~\cite{Bacchetta:2019sam}, MAP22~\cite{Bacchetta:2022awv}, ART23~\cite{Moos:2023yfa}, and HSO24~\cite{Gonzalez-Hernandez:2022ifv}. At short distances, the phenomenological fittings converge and align with the predictions of the ILM, while at large distances the latter remains within the substantial error margins among different phenomenological fits.


\section{Conclusions}
\label{SECV}
We have outlined a general framework for the non-perturbative analysis of soft functions 
in QCD, as they follow from factorization in DY, SIDIS or $e^+e^-$. This analysis uses
the dominant gauge fields revealed by gradient flow cooling of QCD lattice gauge
simulations, as captured by the ILM in Euclidean signature. Key to this analysis is
the evaluation of the parametrization of the soft functions in terms of  Wilson 
loops sloped at angle $\theta$, which is then identified with the soft function rapidity 
through $\theta\rightarrow i\chi$ by analytical continuation~\cite{Meggiolaro:1997mw,Giordano:2008ua,Shuryak:2021hng}. 

We have used this construction to show how the non-perturbative part of the CS kernel receives
significant contribution from the ILM, especially at small resolution 
and large transverse separations. In large resolution, ILM contribution is mixed with perturbative contribution at small transverse separation, but ILM still dominates the large-distance behavior of the CS kernel.

Remarkably, the
weak and strong field instanton analyses reveal almost the same angle dependence of 
the cusp factor, which allows for an almost exact analytical continuation from 
Euclidean to Minkowski signature, which is highly nontrivial. Conversely, current lattice simulations
of the CS kernel can be directly probed versus $h(\theta)$ as developed in this work, and compared to the ILM prediction, making the analytical continuation straightforward. This way provides  for a novel method for the evaluation of the soft function and the CS kernel solely in Euclidean signature, based on the formalism put forth in~\cite{Shuryak:2021hng}.

Similarly, the extracted 
Coulomb self-energy accumulated on the Wilson lines is small and compatible between the weak and strong field  limits. 
Our ILM calculations predict that the asymptotic behavior of the CS kernel at large $ b_\perp $ exhibits logarithmic dependence, offering a crucial constraint for future phenomenological fitting.

To compare our results with the large resolution lattice and phenomenological results,
we have interpolated our evolved ILM at large separation, with the perturbative results 
at small separation. The evolution of the ILM results made use of the reverse gradient 
flow cooling. Our results agree well with the recent lattice results and can extend to larger transverse separations. A better control of the lattice results in this region
would improve the comparison. Our results also compare well to a selected set of phenomenological 
extractions.

\begin{acknowledgments}\noindent
The work of WL and IZ  is supported by the U.S. Department of Energy, Office of Science, Office of Nuclear Physics under Contract No. DE-FG-88ER40388. The work of YZ is supported by the U.S. Department of Energy, Office of Science, Office of Nuclear Physics through Contract No.~DE-AC02-06CH11357. This work is also supported in part by the Quark-Gluon Tomography (QGT) Topical Collaboration, with Award DE-SC0023646.
\end{acknowledgments}

\appendix

\section{Integrals for Wilson loop}
\label{soft_wilson_int}
In the weak field limit,  the instanton and anti-instanton
contributions to the Coulomb self energy and the CS kernel
in the pertinent Wilson loop can be simplified considerably.

\subsection{Coulomb potential in weak field}
In the zero angle limit, the integral for the two parallel Wilson lines can be written as
\begin{widetext}
\begin{equation}
    V^{(1)}_{\rm Coul}=-2\int_{-\infty}^\infty ds\int_{-\infty}^\infty ds'\left(d_1((s-s')^2+b_\perp^2)-d_1((s-s')^2)-\frac{b_\perp^2}{(s-s')^2+b_\perp^2}d_2((s-s')^2+b_\perp^2)\right)
\end{equation}
with manifest translational symmetry. Using the change of variables $s\rightarrow\sigma=s-s'$, the integration over $s'$ is  the length $T$ of the Wilson loop, and the result reads
\begin{equation}
\begin{aligned}
    V^{(1)}_{\rm Coul}=&-4T\int_{0}^\infty d\sigma\left[d_1(\sigma^2+b_\perp^2)-d_1(\sigma^2)-\frac{b_\perp^2}{\sigma^2+b_\perp^2}d_2(\sigma^2+b_\perp^2)\right]\,.
\end{aligned}
\end{equation}
\end{widetext}

\subsection{CS kernel in weak field}
Using  the change of the variables $\sigma=s-s'$ and $\eta=s/\sigma$ and some algebraic simplification, the angle dependence can be factorized out from the rest of the integration
in CS kernel. More specifically, we have 
\begin{widetext}
\begin{equation}
\begin{aligned}
    K^{(1)}(\rho,b_\perp,\theta)=&-4v\cdot\bar v\int_0^1\frac{d\eta}{1-2\eta\bar\eta(1-v\cdot\bar v)}\int_0^\infty d\sigma\sigma d_1[\sigma^2+b_\perp^2]\\
    &+4[1-(v\cdot \bar{v})^2]\int_0^1d\eta\frac{\eta\bar\eta}{[1-2\eta\bar\eta(1-v\cdot\bar v)]^2}\int_0^\infty d\sigma\sigma \frac{\sigma^2}{\sigma^2+b_\perp^2}d_2(\sigma^2+b_\perp^2)\\
    &+4v\cdot\bar{v}\int_0^1\frac{d\eta}{1-2\eta\bar\eta(1-v\cdot\bar v)}\int_0^\infty d\sigma\sigma \frac{b_\perp^2}{\sigma^2+b_\perp^2}d_2(\sigma^2+b_\perp^2)\\
    &+4v\cdot\bar v\int_0^1\frac{d\eta}{1-2\eta\bar\eta(1-v\cdot\bar v)}\int_0^\infty d\sigma\sigma d_1(\sigma^2)\\
    &-4[1-(v\cdot \bar{v})^2]\int_0^1 d\eta \frac{\eta\bar\eta}{[1-2\eta\bar\eta(1-v\cdot\bar v)]^2}\int_0^\infty d\sigma\sigma d_2(\sigma^2)\\
    &-(v\cdot\bar{v}\rightarrow1)\,.
\end{aligned}
\end{equation}
Using the integral identities
\begin{align}
    &\int_0^1\frac{d\eta}{1-2\eta\bar\eta(1-v\cdot\bar v)}=\frac{\theta}{\sin\theta}\,,\nonumber\\
    &\int_0^1d\eta\frac{\eta\bar\eta}{[1-2\eta\bar\eta(1-v\cdot\bar v)]^2}=-\frac{\theta \cot\theta-1}{2\sin^2\theta}\,,
\end{align}
we obtain in the weak field limit
\begin{equation}
\begin{aligned}
    K^{(1)}(\rho,b_\perp,\theta)=&-4h(\theta)\int_0^\infty d\sigma\sigma \left[d_1(\sigma^2+b_\perp^2)-\frac{b_\perp^2}{\sigma^2+b_\perp^2}d_2(\sigma^2+b_\perp^2)-d_1(\sigma^2)\right]\\
    &+2h(\theta)\int_0^\infty d\sigma\sigma \left[\frac{\sigma^2}{\sigma^2+b_\perp^2}d_2(\sigma^2+b_\perp^2)-d_2(\sigma^2)\right]\,.
\end{aligned}
\end{equation}
\end{widetext}

\section{Cusp anomalous dimension}
\label{Appx:cusp}
Here we summarize the cusp anomalous dimension up to next-to-next-to-next-to-leading 
order (N$^3$LO) as~\cite{Korchemsky:1987wg,Moch:2004pa,Henn:2019swt,vonManteuffel:2020vjv} 
\begin{equation}
    \Gamma_{\mathrm{cusp}}(\alpha_s)=\sum_{n=1}^\infty \left(\frac{\alpha_s}{4\pi}\right)^n\Gamma_n\,,
\end{equation}
with the $n$-th order coefficients given by
\begin{equation}
    \Gamma_1=4C_F\,,
\end{equation}
\begin{equation}
\begin{aligned}
    \Gamma_2 =&8 C_F \left[ C_A \left( \frac{67}{18} - \frac{\pi^2}{6} \right) - \frac{5}{9} N_f \right]\\
    =&\frac{1072}{9} - \frac{16 \pi^2}{3} - \frac{160}{27} N_f\,,
\end{aligned}
\end{equation}
\begin{equation}
\begin{aligned}
    &\Gamma_3 = 352 \zeta(3) + \frac{176 \pi^4}{15} - \frac{2144 \pi^2}{9} + 1960 \\
    &+ N_f \left( -\frac{832 \zeta(3)}{9} + \frac{320 \pi^2}{27} - \frac{5104}{27} \right) - \frac{64}{81} N_f^2 \,,
\end{aligned}
\end{equation}
\begin{widetext}
\begin{equation}
\begin{aligned}
    \Gamma_4 =& 
    \Bigg(-1536 \zeta(3)^2 - 704 \pi^2 \zeta(3) + 28032 \zeta(3) - \frac{34496 \zeta(5)}{3} 
    + \frac{337112}{9} - \frac{178240 \pi^2}{27} + \frac{3608 \pi^4}{5} - \frac{32528 \pi^6}{945}\Bigg)\\
&+ N_f \Bigg( 
    \frac{1664 \pi^2 \zeta(3)}{9} - \frac{616640 \zeta(3)}{81} + \frac{25472 \zeta(5)}{9} - \frac{1377380}{243} + \frac{51680 \pi^2}{81} - \frac{2464 \pi^4}{135} 
\Bigg)\\
&+ N_f^2 \Bigg( 
    \frac{16640 \zeta(3)}{81} + \frac{71500}{729} - \frac{1216 \pi^2}{243} - \frac{416 \pi^4}{405} 
\Bigg) + N_f^3 \Bigg( 
    \frac{256 \zeta(3)}{81} - \frac{128}{243} 
\Bigg)\,.\\
\end{aligned}
\end{equation}


\end{widetext}


The perturbative part of the CS kernel at minimal logarithmic scale $K^{(\rm pert)}_{\rm CS}(b_\perp=b_*, \mu=\mu_b)$ has been calculated up to four-loop order~\cite{Li:2016ctv,Vladimirov:2016dll,Moult:2022xzt,Duhr:2022yyp},
\begin{equation}
    K^{(\rm pert)}_{\rm CS}(b_*, \mu_b)=\sum_{n=1}^\infty\left(\frac{\alpha_s(\mu_b)}{4\pi}\right)^nk_n\,.
\end{equation}
 In this work, we use the perturbative CS kernel up to N$^3$LO,
\begin{align}
k_1 =& 0 \,,\\
k_2 =&  8 C_F \left[\left( \frac{7}{2} \zeta(3) - \frac{101}{27} \right) C_A + \frac{14}{27} N_f \right]\nonumber\\
=&-2(-56 \zeta(3) + \frac{1616}{27} - \frac{224}{81} N_f) \,,\\
k_3 =& -2\Bigg( \frac{176 \pi^2 \zeta(3)}{3} - \frac{24656 \zeta(3)}{9} + 1152 \zeta(5) \nonumber\\
&+ \frac{594058}{243} - \frac{6392 \pi^2}{81} - \frac{154 \pi^4}{45} \Bigg)  \nonumber\\
&-2 N_f \left( \frac{7856 \zeta(3)}{81} - \frac{166316}{729} + \frac{824 \pi^2}{243} + \frac{4 \pi^4}{405} \right) \nonumber\\
& -2 N_f^2 \left( \frac{64 \zeta(3)}{27} + \frac{3712}{2187} \right)\,.
\end{align}

\bibliography{ref}

\end{document}